\RequirePackage{lineno}
\documentclass[aps,twocolumn,showpacs,byrevtex]{revtex4-1}
\usepackage{times}
\usepackage{amsmath}
\usepackage{graphicx}
\usepackage{color}
\usepackage[T1]{fontenc}
\usepackage{bm,color}
\usepackage{float}
\usepackage[colorlinks,
            linkcolor=blue,
            citecolor=blue,
            urlcolor=blue,
            ]{hyperref}
\newcommand{\y}{\psi(4260)}
\newcommand{\yy}{\psi(4360)}
\newcommand{\yyy}{\psi(4660)}
\newcommand{\z}{Z_c(3900)}
\newcommand{\x}{\psi_2(3823)}
\newcommand{\pp}{\pi^+\pi^-}

\newcommand{\LL}{\ell^+\ell^-}
\newcommand{\EE}{e^+e^-}
\newcommand{\MM}{\mu^+\mu^-}
\newcommand{\GG}{\gamma\gamma}
\newcommand{\etap}{\eta^\prime}
\newcommand{\psip}{\psi(2S)}

\newcommand{\jpsi}{J/\psi}
\newcommand{\piz}{\pi^0}
\newcommand{\chico}{\chi_{c1}}
\newcommand{\chict}{\chi_{c2}}

\newcommand{\ppjpsi}{\pi^+\pi^-J/\psi}

\def\Journal#1#2#3#4{{#1} {\bf #2}, #3 (#4)}

\def\PRL{Phys. Rev. Lett.}
\def\PRD{Phys. Rev. D}

\begin{document}

\title{\boldmath Observation of Resonance Structures in $e^+e^-\to \pi^+\pi^-\psi_2(3823)$ and Mass Measurement of $\psi_2(3823)$
}

\author{
\begin{small}
\begin{center}
M.~Ablikim$^{1}$, M.~N.~Achasov$^{10,b}$, P.~Adlarson$^{69}$, M.~Albrecht$^{4}$, R.~Aliberti$^{29}$, A.~Amoroso$^{68A,68C}$, M.~R.~An$^{33}$, Q.~An$^{65,51}$, X.~H.~Bai$^{59}$, Y.~Bai$^{50}$
% ...
\\
\vspace{0.2cm}
(BESIII Collaboration)\\
\vspace{0.2cm} {\it
$^{1}$ Institute of High Energy Physics, Beijing 100049, People's Republic of China\\
% ...
$^{4}$ Bochum Ruhr-University, D-44780 Bochum, Germany\\
% ...
$^{10}$ G.I. Budker Institute of Nuclear Physics SB RAS (BINP), Novosibirsk 630090, Russia\\
% ...
$^{29}$ Johannes Gutenberg University of Mainz, Johann-Joachim-Becher-Weg 45, D-55099 Mainz, Germany\\
% ...
$^{33}$ Liaoning Normal University, Dalian 116029, People's Republic of China\\
% ...
$^{50}$ Southeast University, Nanjing 211100, People's Republic of China\\
$^{51}$ State Key Laboratory of Particle Detection and Electronics, Beijing 100049, Hefei 230026, People's Republic of China\\
% ...
$^{59}$ University of Jinan, Jinan 250022, People's Republic of China\\
% ...
$^{65}$ University of Science and Technology of China, Hefei 230026, People's Republic of China\\
% ...
$^{68}$ University of Turin and INFN, (A)University of Turin, I-10125, Turin, Italy; (B)University of Eastern Piedmont, I-15121, Alessandria, Italy; (C)INFN, I-10125, Turin, Italy\\
$^{69}$ Uppsala University, Box 516, SE-75120 Uppsala, Sweden\\
% ...
$^{b}$ Also at the Novosibirsk State University, Novosibirsk, 630090, Russia\\
% ...
}\end{center}
\vspace{0.4cm}
\end{small}
}

\date{\today}

\begin{abstract}

  Using a data sample corresponding to an integrated luminosity of 11.3 $\rm fb^{-1}$ collected at center-of-mass energies from
  $4.23$ to $4.70$~GeV with the BESIII detector, we measure the product of
  the $e^+e^-\to \pi^+\pi^-\psi_2(3823)$ cross section and the branching fraction $\mathcal{B}[\psi_2(3823)\to \gamma\chi_{c1}]$.
  For the first time, resonance structure is observed in the cross section line shape of $e^+e^-\to \pi^+\pi^-\psi_2(3823)$ with significances
  exceeding $5\sigma$. 
  A fit to data with two coherent Breit-Wigner resonances modeling the $\sqrt{s}$-dependent cross section yields
  $M(R_1)=4406.9\pm 17.2\pm 4.5$~MeV/$c^2$, $\Gamma(R_1)=128.1\pm 37.2\pm 2.3$~MeV,
  and $M(R_2)=4647.9\pm 8.6\pm 0.8$~MeV/$c^2$, $\Gamma(R_2)=33.1\pm 18.6\pm 4.1$~MeV.
  Though weakly disfavored by the data, a single resonance with $M(R)=4417.5\pm26.2\pm3.5$~MeV/$c^2$,
  $\Gamma(R)=245\pm48\pm13$~MeV is also possible to interpret data.
  This observation deepens our understanding of the nature of the vector charmoniumlike states.
  The mass of the $\psi_2(3823)$ state is measured as $(3823.12\pm 0.43\pm 0.13)$~MeV/$c^2$, 
  which is the most precise measurement to date. 
\end{abstract}


\maketitle
In the quark model, hadrons are strongly-interacting, composite particles 
built from colour-neutral combinations of quarks and antiquarks~\cite{quarkmodel}.
It was long thought that all observed hadrons fall
into two classes only: baryons, composed of three quarks,
and mesons, bound states of a quark--antiquark pair.
The 
QCD theory describing the strong interaction
also allowed for
other colour-neutral configurations, but there was no experimental evidence for
such `exotic' hadrons.

This simple picture, however, has been challenged since 2003, 
when many new charmonium-like states 
such as the 
$\chi_{c1}(3872)$~\cite{belle-x3872}, $\y$~\cite{babar-y4260}, and $\z$~\cite{bes3-zc3900,belle-zc3900} 
have been observed experimentally. These 
particles 
can not easily be accommodated in the spectrum of conventional charmonium states and are widely 
considered to be promising candidates 
for QCD exotic hadrons~\cite{rev-olsen, rev-qwg}.
Among them, the vector $\psi$-states usually couple to hidden charm final states like $J/\psi$, $\psi(2S)$, or $h_c$
via dipion transitions, such as the $\y\to\ppjpsi$~\cite{babar-y4260,belle-y4260,bes3-y4260}, 
$\yy/\yyy\to\pp\psip$~\cite{belle-y4360,babar-y4360}, and $\psi(4390)\to\pp h_c$~\cite{bes3-y4390}.
In addition, there are also vector states with mass above 4.6~GeV reported in 
$\EE\to\Lambda_c^+\Lambda_c^-$~\cite{belle-y4630},
and $\EE\to D_s^+D_{s1}(2536)^-/D_s^+D^*_{s2}(2573)^-$ processes~\cite{belle-y4626}.
At the moment, experimental information about these $\psi$-states, 
especially for the high mass states is still quite limited. 
It is not clear whether $\yy$ and $\psi(4390)$ correspond to the same resonance or not. 
Above 4.6~GeV, the resonance parameters of vector states observed in hidden-charm and open-charm
final states are not exactly the same. Whether there exists one or more resonances 
is a long-standing puzzle in the study of the vector charmonium-like spectrum.
To pin down these issues, new observations from
experiment are urgently needed.

One of the vector states, the $\yyy$ resonance, was first observed
by the Belle experiment~\cite{belle-y4360} and later confirmed by the $BABAR$
and BESIII experiments~\cite{babar-y4660,bes3-y4360}.
It remains, however, unclear what the exact nature is of the $\yyy$. 
Possible interpretations of its internal structure 
include a hadronic molecule~\cite{f0-psip},
a baryonium~\cite{qcf}, or a compact tetraquark state~\cite{tetra}.
For these theoretical models, the coupling of $\yyy$ to
the $\psip$ state with no or weak coupling to other charmonium states 
is an essential element. Therefore, a search for the decay of
$\yyy$ to final state other than $\psip$ in experiment
helps to test various pictures for the $\yyy$ structure.

The $D$-wave charmonium state $\x$~\cite{psi2,full-rec} and $\psi_3(3842)$~\cite{psi3}
were well established recently, and several decay modes of the $\x$ state are also observed~\cite{psi2-decay}.
It serves as a new probe to
study the vector $\psi$-states. We search for the dipion transition
of $\yyy$ to $\x$, which on the one hand helps
establish the $\yyy$ state, on the other hand sheds light on
its internal structure. At the same time, 
the $\x$ mass
is also precisely measured, which can be used to calibrate the parameters
in the potential model~\cite{potential2}, 
and finally greatly deepens our understanding
of the dynamics of the $c\bar{c}$ system.

In this Letter, we measure the $\sqrt{s}$-dependent production cross section of the process
$\EE\to\pp\x$, and explore the resonance structures in the cross section line shape.
The resonance parameters of the $\x$ state are measured as well.
To increase the yield of signal events, a partial reconstruction approach is employed.
We use a data sample corresponding to an integrated luminosity of 11.3 $\rm fb^{-1}$, taken at center-of-mass (c.m.) energies from $\sqrt{s}=4.23$ to $4.70$~GeV~\cite{lum},
with the BESIII detector~\cite{bes3-detector} operating at the BEPCII storage ring~\cite{Yu:IPAC2016-TUYA01}.
The $\x$ candidates are reconstructed in their $\gamma\chico$
decay mode, with $\chi_{c1}\to \gamma J/\psi$
and $J/\psi\to \LL$ ($\ell=e$ or $\mu$).

The BESIII detector is described in detail elsewhere~\cite{bes3-detector,etof}.
A {\sc geant4}-based~\cite{geant4} Monte Carlo~(MC) simulation 
software package 
is used to optimize event selection
criteria, determine the detection efficiency, and estimate the
backgrounds. For the signal process, we generate 50,000 $\EE\to \pp\x$ 
events at each c.m.\ energy
using an {\sc evtgen}~\cite{evtgen} phase space model.
Initial-state-radiation (ISR) is simulated with {\sc kkmc}~\cite{kkmc}, where we use the
$\EE\to\pp\x$ cross section measured from this analysis as input.
The maximum ISR photon energy is set to correspond
to the production threshold of the $\pp\x$ system at 4.1~GeV/$c^2$.
Final-state-radiation is simulated with {\sc photos}~\cite{photos}.


Events with four good charged tracks with net charge zero are selected
as described in Ref.~\cite{full-rec}.
Electromagnetic showers identified as photon candidates must satisfy fiducial
shower quality as well as timing requirements as described in Ref.~\cite{bes3_hc}. 
For events with only one photon candidate ($N_{\gamma}=1$),
assuming that only one of the two radiative photons is detected, 
we use a partial reconstruction strategy, 
\emph{i.e.} allowing one missing photon ($\gamma_{\rm miss}$). 
The mass square of the
missing photon candidate is required to be $-0.07<M^2_{\rm miss}(\gamma)<0.08$~GeV$^2$/$c^4$
(with a signal efficiency $>99\%$),
where the 4-momentum of $\gamma_{\rm miss}$ is computed from momentum conservation.
To improve the momentum and energy resolution and to further reduce
background, a one-constraint (1C) kinematic fit is performed under the
hypothesis of $\pp\LL\gamma\gamma_{\rm miss}$ 
to the initial $\EE$ c.m.\ system. 
The $\chi^2/{\rm ndf}$ of the kinematic fit is required to
be less than $15/1$. For multi-photon events ($N_{\gamma} \ge 2$),
we use the full reconstruction strategy as described in Ref.~\cite{full-rec}.

To reject radiative Bhabha and radiative dimuon
($\gamma\EE/\gamma\MM$) background events with
gamma-conversion ($\gamma\to\EE$), where the converted electrons are 
misidentified as pions, the cosine of the opening angle of the pion pairs
is required to be less than 0.98. The background from
$\EE\to \eta\jpsi$ with $\eta\to \pp\pi^0/\gamma\pp$ is effectively rejected
by the invariant mass requirement $M(\GG_{\rm miss}\pp)>0.65~{\rm GeV}/c^2$.
In order to remove possible backgrounds from $\EE\to\gamma_{\rm ISR}\psip$, 
$\EE\to\eta\psip$ with $\eta\to\GG$, and $\EE\to\GG\psip$ processes, 
the invariant mass of $\ppjpsi$ is required to
satisfy $|M(\ppjpsi)-m[\psip]|>7$~MeV/$c^2$~\cite{Mppjpsi}.

According to a resolution of $(14.2\pm0.5)$~MeV from $\psip$ data events for the $M(\LL)$ mass,
the $\jpsi$ signal region is defined as $3.06<M(\LL)<3.135$~GeV/$c^2$. 
To estimate non-$\jpsi$ backgrounds, we also define 
$\jpsi$ mass sidebands as $2.950<M(\LL)<3.025$~GeV/$c^2$ or
$3.170<M(\LL)<3.245$~GeV/$c^2$. 
To reconstruct the $\chi_{c1}$ candidate from the $\x$ decay, 
the 4-momenta of the two radiative photons after the 1C kinematic fit are boosted to the c.m.\ frame of the $\x$ system. 
The photon with the higher energy is used to reconstruct $\chico$, 
while the lower-energy one is considered to originate from the $\x$ decay.
MC simulation shows that the mis-assignment 
of the two photons is negligibly small ($<1\%$).
The mass window of the reconstructed $\chico$ candidates is defined as
$3.48<M(\gamma_H\jpsi)<3.53$~GeV/$c^2$~\cite{Mppjpsi}, with a signal efficiency of 96\%.

The possible remaining backgrounds mainly come from
$\EE\to(\etap/\gamma\omega)\jpsi$, with $(\etap/\omega)\to
\GG\pp/\gamma\pp$, and $\pp\pp(\piz/\GG)$. 
The $\EE\to(\etap/\gamma\omega)\jpsi$
backgrounds are measured by BESIII using the same
data set~\cite{etap-jpsi, gam-w-jpsi} and can be reliably 
simulated. The $\EE\to \pp\pp(\piz/\GG)$ continuum
background can be estimated by data in the $\jpsi$ mass sidebands.
All these background sources are found to be small,
and only produce flat distributions in the $\x$ signal region. 

\begin{figure}
\begin{center}
\includegraphics[height=1.2in]{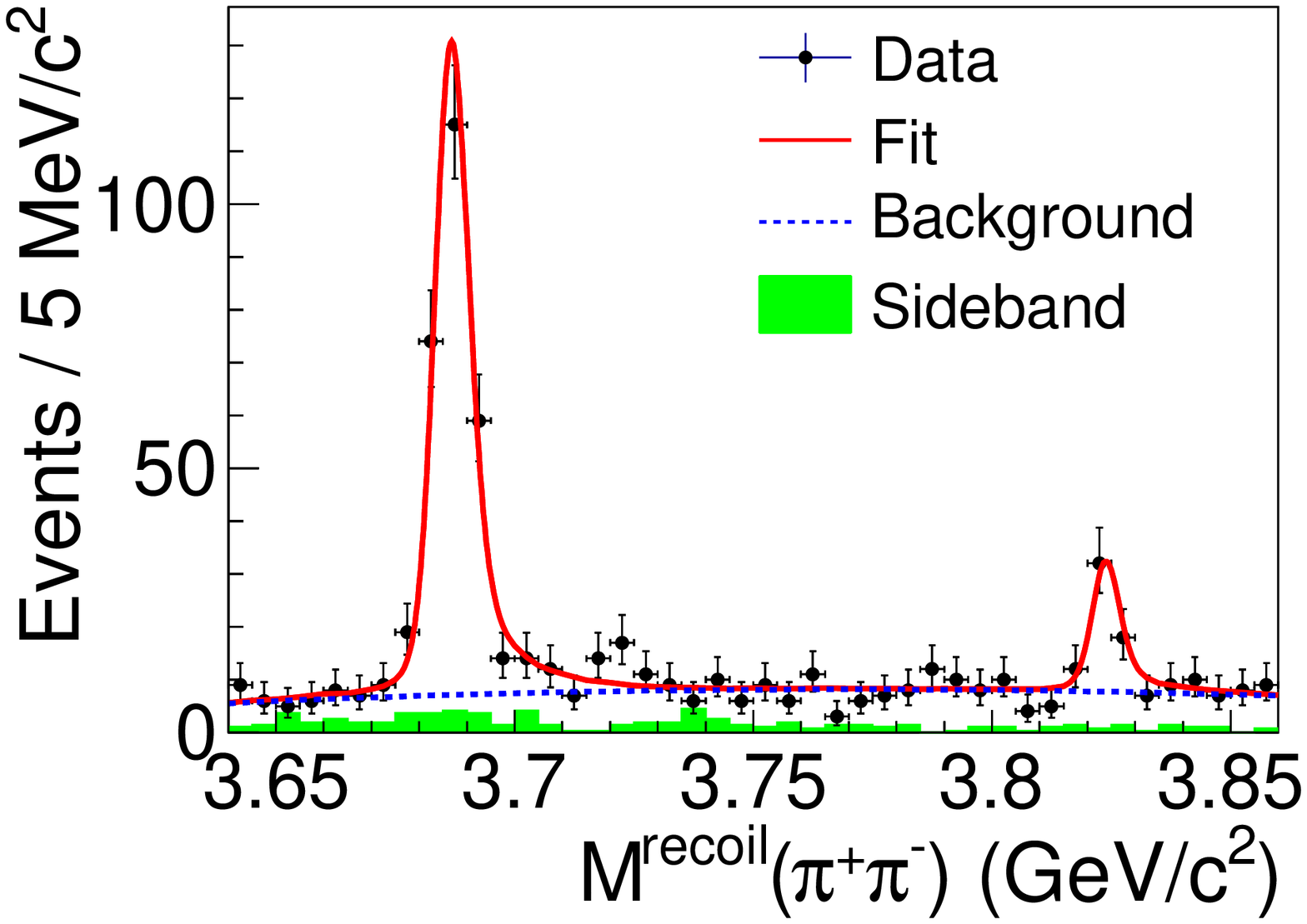}
\includegraphics[height=1.2in]{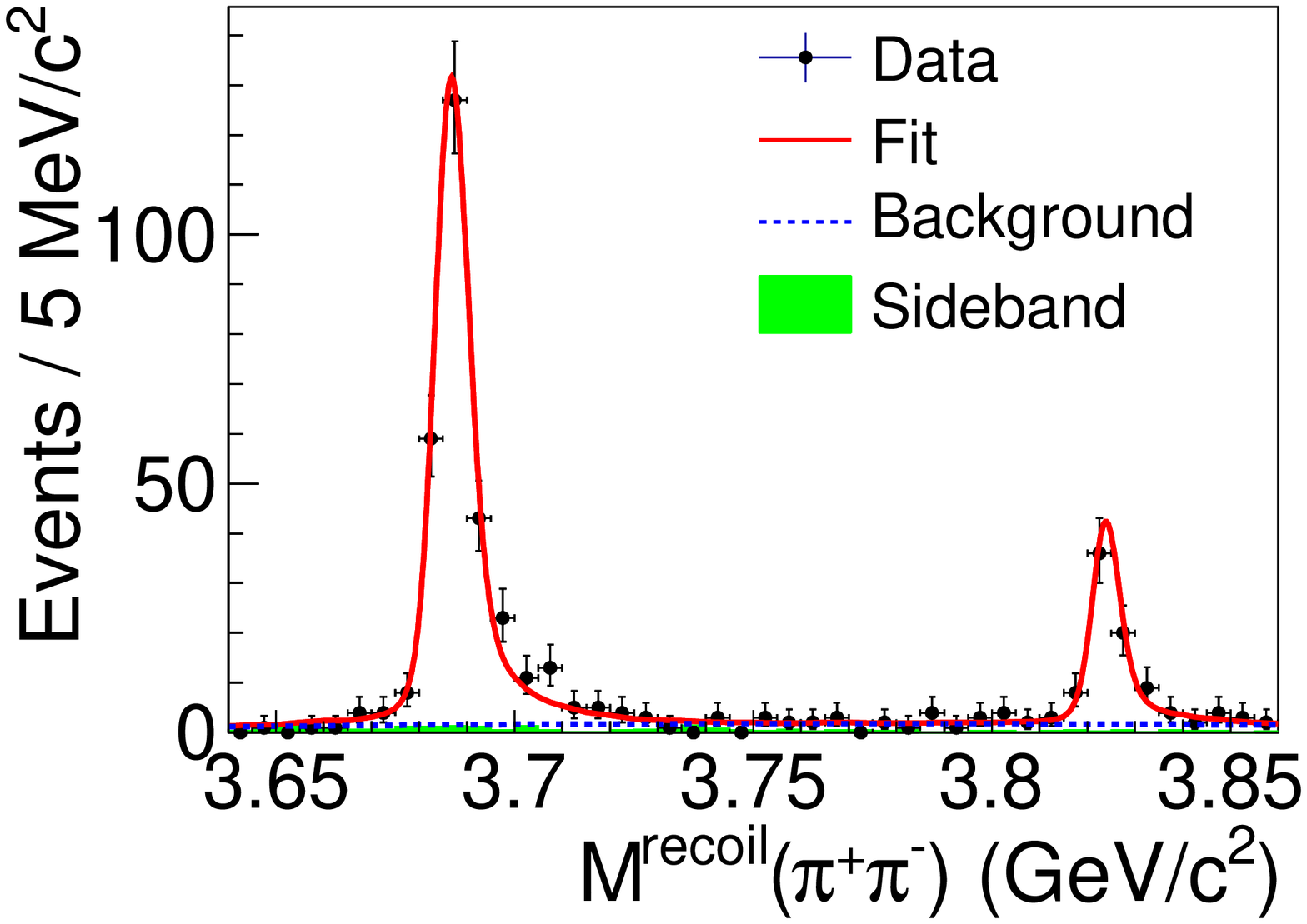}
\caption{Result of the simultaneous fit to the $M^{\rm recoil}(\pp)$ distributions 
for one-photon events (left) and multi-photon events (right).
Dots with error bars are the selected data, the red solid curves are fit results, the blue
dashed curves are backgrounds, and the green shaded
histograms are backgrounds estimated from $\jpsi$ mass sideband events.}
\label{part-full}
\end{center}
\end{figure}

To achieve better sensitivity, the one-photon events (partial reconstruction) and the multi-photon events (full reconstruction) are separated. 
Figure~\ref{part-full} shows the $M^{\rm recoil}(\pp)$ distributions for data, 
where obvious $\psip$ and $\x$ signal peaks are observed in both the one-photon and multi-photon events.
Here, $M^{\rm recoil}(\pp)=\sqrt{(P_{\EE}-P_{\pi^+}-P_{\pi^-})^2}$ is the recoil mass of $\pp$, where $P_{\EE}$ and $P_{\pi^\pm}$ are the 4-momenta of the initial $\EE$ system and the reconstructed $\pi^\pm$ candidates, respectively. 
For this expression, we use the $\pi^\pm$ momenta without the kinematic fit correction because of the good resolution for low momentum pions according to MC simulation studies.
A simultaneous unbinned extended maximum likelihood fit to the two $M^{\rm recoil}(\pp)$ distributions is performed
to determine the parameters of the $\x$ state.
In the fit, the signal probability density function (PDF) is represented by $\psip$ and $\x$ 
(with input mass of $3.823$~GeV/$c^2$ and a zero natural width) MC simulated shapes, 
convolved with Gaussian functions with free mean $\mu$ and width $\sigma$
to account for the mass and resolution difference between data and MC simulation, respectively. 
The background shape is parameterized as a second-order polynomial. 

The fit results, also shown in Fig.~\ref{part-full}, yield
$M[\x]=M[\x]_{\rm input}+\mu_{\x}-\mu_{\psip} = 3823.12\pm
0.43$~MeV/$c^2$, where $M[\x]_{\rm input}$
is the input $\x$ mass in MC simulation; $\mu_{\x}=1.02\pm0.43$~MeV/$c^2$ and
$\mu_{\psip}=0.90\pm0.22$~MeV/$c^2$ are the mass shift values for the $\x$ and $\psip$
shapes, respectively.
The total number of $\x$ signal events determined from the fit is $120.0\pm13.6$. The statistical significance of
the $\x$ signal is estimated to be $13.4\sigma$, by
comparing the difference between the log-likelihood value
[$\Delta(\ln\mathcal{L})=96.6$] with or without the $\x$ signal
in the fit and taking the change of the number of degrees of freedom
($\Delta {\rm ndf}=4$) into account.
We are not able to measure the intrinsic width of $\x$ precisely because of the limited data sample size. 
From a fit using a Breit-Wigner (BW) function (with a width parameter that is left free) convolved with a double Gaussian function as signal PDF for $\x$, 
we set an upper limit of $\Gamma[\x]<2.9$~MeV at the $90\%$ confidence level (C.L.).


The product of the $\sqrt{s}$-dependent $\EE\to\pp\x$ cross section and the branching
ratio of $\x\to\gamma\chico$ is calculated as
$\sigma[\EE\to\pp\x]\cdot\mathcal{B}[\x\to\gamma\chico]=\frac{N^{\rm sig}} {\mathcal{L}_{\rm int}
\epsilon \mathcal{B} (1+\delta)}$, where
$N^{\rm sig}$ is the number of $\x\to\gamma\chi_{c1}$ signal events 
obtained from a same fit ($\sigma$ fixed to previous result) to the $M^{\rm recoil}(\pp)$ distribution at a certain c.m. energy,
$\mathcal{L}_{\rm int}$ is the integrated luminosity, $\epsilon$ is the detection
efficiency, $\mathcal{B}$ is the branching fraction of
$\chico\to \gamma\jpsi\to \gamma \LL$, and ($1+\delta$) is
the radiative correction factor, which depends on the cross section 
line shape of $\EE\to \pp\x$. Since visible enhancements are observed
near $4.40$ and $4.65$~GeV in the cross section line shape,
the radiative correction factors
are first obtained by modelling the line shape with two coherent BW resonances, and
then iterated by updating the cross section measurement until this procedure converges,
with a relative difference for $(1+\delta)\epsilon<1\%$ between the last two iterations.
The numerical results of the cross section measurement are listed in 
the supplemental material~\cite{supplement}.

\begin{figure}
\begin{center}
\includegraphics[height=1.2in]{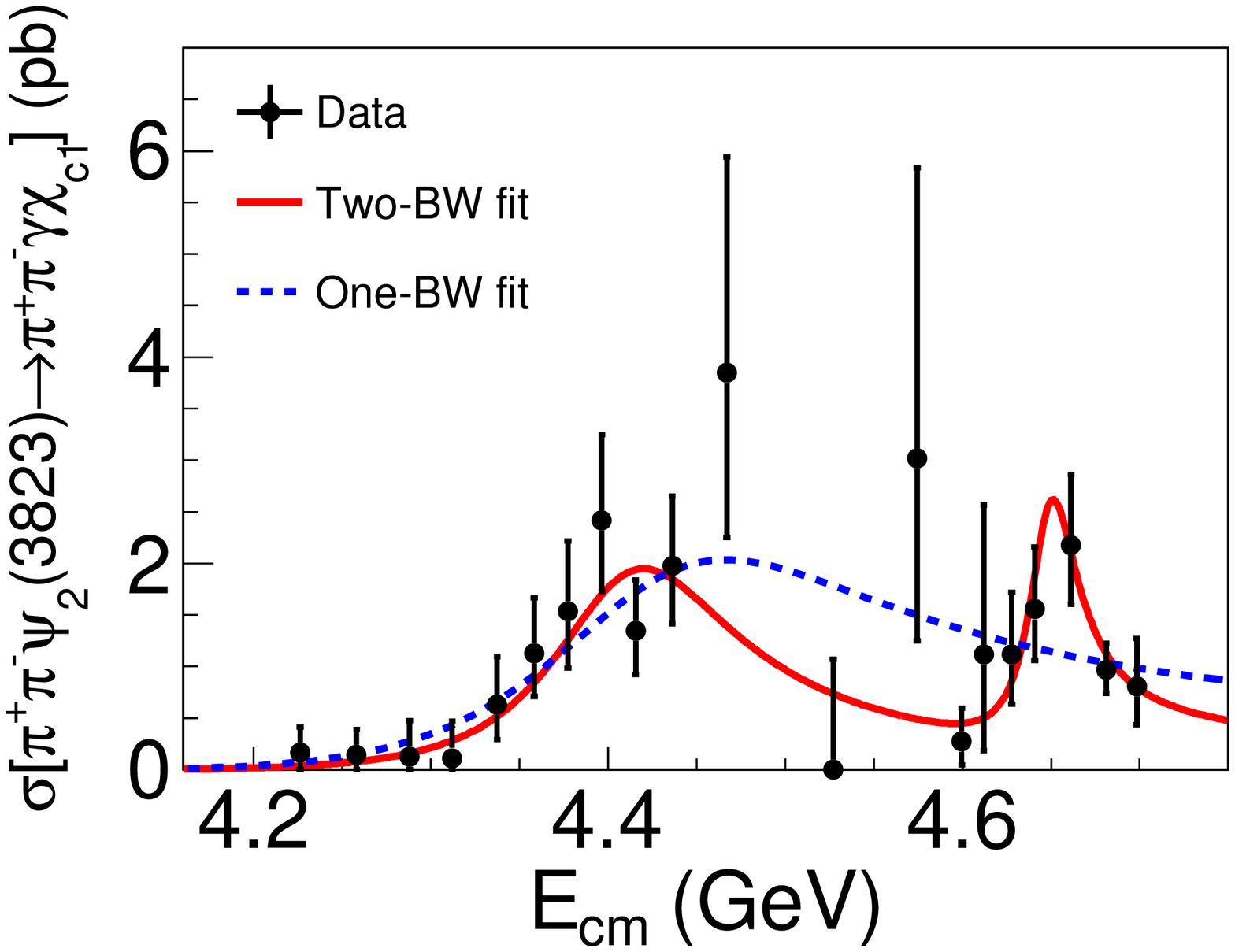}
\includegraphics[height=1.2in]{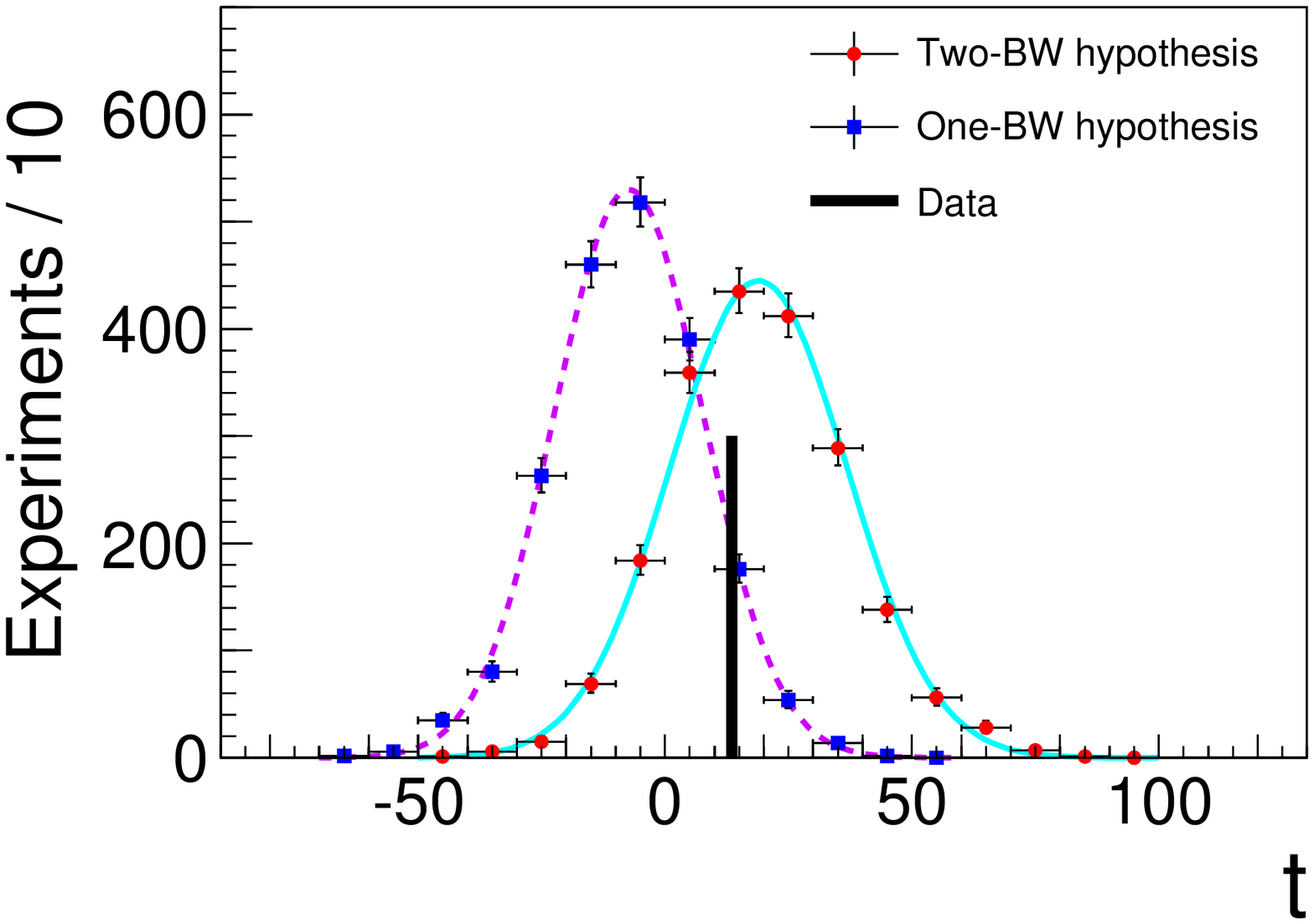}
\caption{(Left panel) result of the fit to the $\sqrt{s}$-dependent cross section 
$\sigma[\EE\to\pp\x]$ times the branching ratio $\mathcal{B}[\x\to\gamma\chico]$.
Dots with error bars are data, and the red solid (blue dashed) curve 
shows the fit with two coherent resonances (one resonance); 
(right panel) the likelihood ratio $t=-2\ln\frac{\mathcal{L}_{\rm 1BW}}{\mathcal{L}_{\rm 2BW}}$
distribution from MC pseudo-experiments under two cross section hypothesis. Red dots (blue squares) with
error bars are the two resonances (one resonance) hypothesis, and the black line shows data measurement.}
\label{xs}
\end{center}
\end{figure}

To extract the resonance structures in $\sigma[\EE\to\pp\x]$, 
a maximum likelihood fit using the coherent sum of two BW resonances
to model the measured cross section is performed to data events in the
$\x$ signal interval [3.815,3.835]~GeV/$c^2$. The likelihood is constructed as that in Ref.~\cite{likelihood}.
There are two solutions with identical fit quality, and 
all resonance parameters from the fit are summarized in Table~\ref{res-par}.
In addition, a fit with one single BW resonance to model the cross section
yields $M[R]=4417.5\pm26.2$~MeV/$c^2$, 
$\Gamma_{\rm tot}[R]=245\pm48$~MeV, 
$\Gamma_{\rm e^+e^-} \mathcal{B}_{1} \mathcal{B}_{2}=0.57\pm0.08$~eV/$c^2$.
The fit result is shown in Fig.~\ref{xs}. 
To discriminate the two resonances hypothesis ($H_1$) from the one resonance hypothesis ($H_0$)
for the cross section interpretation, the likelihood ratio $t=-2\ln\frac{\mathcal{L}_{\rm 1BW}}{\mathcal{L}_{\rm 2BW}}$ 
is used as a test variable. We perform 2000 MC pseudo-experiments for both hypotheses and
the corresponding $t$-distributions are shown in Fig.~\ref{xs}. The $t=13.6$ from data is positive
and slightly favors $H_1$. The $p$-value to reject $H_0$ is 8.2\%, corresponding
to a significance of $1.7\sigma$.
Other possible continuum parametrizations of the cross section in the fit, such as 
a shape of three-body phase space, $1/s^{n}$, or a product of phase space with $1/s^{n}$
are also tested, and they are not able to describe data well.
The significance for the resonance hypothesis (with either one or two resonances)
over continuum is estimated to be greater than $5\sigma$.

\begin{table}
\begin{center}
\caption{Results of the fit to the distribution of  $\sigma[\EE\to\pp\x]\cdot\mathcal{B}[\x\to\gamma\chico]$ with two coherent resonances. Here,
$M[R_i]$ and $\Gamma_{\rm tot}[R_i]$ represent the mass (in MeV/$c^2$) and total width (in MeV) of resonance $R_i$, respectively; 
$\Gamma_{\rm e^+e^-} \mathcal{B}_{1}^{R_i} \mathcal{B}_{2}$ is the product of the $\EE$ partial width (in eV/$c^2$) 
and branching fraction of $R_i\to\pp\x\to\pp\gamma\chico$ ($i=1$, $2$). 
The parameter $\phi$ (in degrees) is the relative phase between the two resonances. 
The first uncertainties are statistical and the second systematic.} 
\label{res-par}
\begin{tabular}{cc}
\hline\hline
Parameters & Solution I~~~~~~~~~~Solution II \\
\hline
$M[R_1]$ & $4406.9\pm17.2\pm4.5$ \\
$\Gamma_{\rm tot}[R_1]$ & $128.1\pm37.2\pm2.3$ \\
$\Gamma_{\rm e^+e^-} \mathcal{B}_{1}^{R_1} \mathcal{B}_{2}$ & $0.36\pm0.10\pm0.03$~~~~~$0.30\pm0.09\pm0.03$ \\
$M[R_2]$ & $4647.9\pm8.6\pm0.8$ \\
$\Gamma_{\rm tot}[R_2]$ & $33.1\pm18.6\pm4.1$ \\
$\Gamma_{\rm e^+e^-} \mathcal{B}_{1}^{R_2} \mathcal{B}_{2}$ & $0.24\pm0.07\pm0.02$~~~~~$0.06\pm0.03\pm0.01$ \\
$\phi$ & $267.1\pm16.2\pm3.2$~~~~$-324.8\pm43.0\pm5.7$ \\
\hline\hline
\end{tabular}
\end{center}
\end{table}


The systematic uncertainties in the $\x$ mass measurement include
those from the absolute mass scale, resolution,
parameterization of the $\x$ signal and background shapes. 
%
In the $\x$ mass measurement, we use the $\psip$ mass to calibrate
the absolute mass scale. The uncertainty 
from the $\psip$ mass measurement is therefore taken as the systematic 
uncertainty due to the absolute mass scale, which is $0.12$~MeV/$c^2$. 
To increase the $\psip$ sample size and thus reduce the $\psip$ mass uncertainty, 
we also employ $\psip\to\gamma\chict$ and $\psip\to\eta\jpsi$ data events.
%
The resolution difference between data and MC simulation is also estimated using the $\psip$ events.
Fixing the resolution from a free value to the one measured with $\psip$ events,
the mass difference for $\x$ in the fit is $0.01$~MeV/$c^2$. 
%
In the nominal fit, the signal PDF of $\x$ is parameterized as a MC simulated shape convolved with Gaussian resolution. 
A signal PDF parameterized as a BW convolved with Gaussian resolution is also tested, 
and the mass difference 
($0.03$~MeV/$c^2$) is taken as the systematic uncertainty from signal parameterization.
%
Changing the background shape from a second-order 
polynomial to a linear term yields $0.03$~MeV/$c^2$ mass difference
associated with the background shape parameterization. 
%
Assuming that all the sources are independent, the total systematic
uncertainty is calculated by adding them in quadrature, 
resulting in $0.13$~MeV/$c^2$ for the $\x$ mass measurement. 
For the $\x$ width, 
we measure the upper limits with all of the above systematic uncertainty sources, and
report the most conservative one.


The systematic uncertainties in the cross section measurement
mainly come from luminosity measurement, efficiencies, kinematic fit, signal shape, background
shape, decay model, radiative correction, branching ratios and MC sample size. 
The luminosity is measured using Bhabha events, with an uncertainty of $1.0\%$~\cite{lum}. 
The uncertainty in the tracking efficiency for high momentum
leptons is $1.0\%$ per track.  Pions have momenta between $0.1$ and $0.6$~GeV/$c$,
and the momentum-weighted uncertainty is $1.0\%$ per track. 
By requiring at least one good photon candidate to be detected, 
the photon detection efficiency is very high and the uncertainty is negligible.
The systematic uncertainty for the choice of $\jpsi$ mass window is similar to that of Ref.~\cite{x3872},
which is $1.6\%$.
A track helix parameters correction method as discussed in Ref.~\cite{kf-correction} 
is applied to each MC simulated event during the 1C kinematic fit. The difference in detection 
efficiencies with or without corrections, $1.7\%$, is assigned as the systematic uncertainty from kinematic fit. 
The same sources of signal and background shape parameterizations as discussed for
the $\x$ mass measurement would contribute $3.9\%$ and $1.4\%$ differences in
the $\x$ signal events yields, which are taken as systematic
uncertainties in the cross section measurement.
We model the $\EE\to\pp\x$ process with $L=2$ between $\pp$ and $\x$ in the MC
simulation. The efficiency difference between this model and a
three-body phase space model is $1.8\%$.
For the radiative correction, we take an alternative cross section
line shape from one BW resonance model, and the difference for $(1+\delta)\epsilon$ 
to the nominal two BW resonances model is $5.0\%$.
The uncertainties on the branching ratios for $\chico\to
\gamma\jpsi$ ($2.9\%$) and $\jpsi\to \LL$ ($0.5\%$) are taken from PDG~\cite{pdg}.
The uncertainty from MC sample size is $0.6\%$.
Assuming that all the sources are independent, the total systematic
uncertainty is calculated by adding them in quadrature, 
resulting in $8.8\%$ for the cross section measurement.


The systematic uncertainties for the resonance parameters in the
cross section fit come from absolute c.m. energy measurement, 
the cross section uncertainty, and the fit model.
The c.m. energies of data sets taken in different time periods are measured
with different methods. Shifting the c.m. energies of data sets 
taken in the same period globally (\emph{i.~e.}~fully correlated) within uncertainties, 
we repeat the cross section fit. The deviations of the resonance parameters
are taken as systematic uncertainties. 
%
The systematic uncertainties on the cross section measurements 
are common to all c.m. energies and are propagated to
$\Gamma_{\rm e^+e^-} \mathcal{B}_{1} \mathcal{B}_{2}$ with the same
amount. We quote $8.8\%$ systematic uncertainty for 
$\Gamma_{\rm e^+e^-} \mathcal{B}_{1} \mathcal{B}_{2}$.
BW functions with constant full widths are used as the PDF in the cross section fit. 
We also use BW functions with $\sqrt{s}$-dependent full widths
as the fit PDF, and
the deviations of the resonance parameters between this fit and 
the nominal one are taken as systematic uncertainties from fit model.
%
All these systematic contributions are listed in the supplemental material~\cite{supplement}.
Assuming all the sources are independent, 
the total systematic uncertainties are calculated by adding them in quadrature.


In summary, the product of the $\EE\to\pp\x$ cross section and the branching ratio of
$\x\to\gamma\chico$, 
is measured with 
11.3~fb$^{-1}$ data collected with the BESIII detector at $\sqrt{s}=4.23$ to $4.70$~GeV.
For the first time, we observe resonance structure in the cross section line shape 
with a significance greater than $5\sigma$.
A fit to data with a sum of two coherent BW resonances to model the cross section
yields the masses and widths of both resonances as shown in Tab.~\ref{res-par}. 
Although
weakly disfavored by data with $1.7\sigma$, a single resonance with a mass
$4417.5\pm26.2\pm3.5$~MeV/$c^2$, and a width $245\pm48\pm13$~MeV is also possible to
interpret data. Such a resonance has not been observed before.
This is the first observation of vector $\psi$-states decaying to $D$-wave charmonium state, 
which provides new insights about the $\psi$-states wave functions.
Considering that the measured
$\EE\to\pp\psi(3770)$ cross section is also relatively large near 4.4~GeV~\cite{bes3-3d1}, 
this indicates that the coupling between the $\psi$-states and $D$-wave charmonium might be popular, which
should be taken into account when explaining the nature of these $\psi$-states.

Within current uncertainties, the parameters of structures in the two resonances interpretation are similar to
the $\yy$ and $\yyy$ states reported in $\pp\psip$~\cite{belle-y4360,babar-y4360}.
Assuming the observed structures correspond to these resonances, 
this will be the second decay channel of the mysterious $\yyy$ state after more than 15 years of discovery.
By comparing the measured cross section of $\sigma[\EE\to\pp\x]$ and $\sigma[\EE\to\pp\psip]$~\cite{bes3-y4360},
we find $\frac{\mathcal{B}[\yyy\to\pp\x]\cdot\mathcal{B}[\x\to\gamma\chico]}{\mathcal{B}[\yyy\to\pp\psip]}$
reaches 10\% level.
Taking the branching fraction of $\mathcal{B}[\x\to\gamma\chico]\sim 50\%$~\cite{3d2-br} as input, 
we obtain the relative partial decay width 
$\frac{\Gamma[\yyy\to\pp\x]}{\Gamma[\yyy\to\pp\psip]}\sim 20\%$.
This sizeable partial width poses a challenge to the $f_0(980)\psip$ hadron molecule
interpretation~\cite{f0-psip} for the $\yyy$ nature, which expects $\yyy$
predominantly decaying into $f_0(980)\psip$. The observed $\yyy\to\pp\x$ decay
also differs from an extended baryonium picture~\cite{qcf} which explains the 
$\yyy$ as a $\Sigma_c^0\bar{\Sigma_c^0}$ baryonium and speculates 
$\yyy$ is a first radial excitation in accordance with the $n=2$ radial quantum number
of $\psip$ and absent coupling to charmonium states with $n=1$.
A similar argument also appears in a diquark-antidiquark tetraquark explanation~\cite{tetra},
which assigns the $\yyy$ as the radial excitation of the $\y$ (a $P$-wave tetraquark)
based on the only observed decay $\yyy\to\pp\psip$.
Our observation obviously deviates from this assignment.

We also measure the mass of the $\x$ state as $M[\x]=3823.12\pm0.43\pm0.13~{\rm MeV}/c^2$,
where the first uncertainty is statistical and the second systematic.
The $\x$ width is studied, and an upper limit 
$\Gamma[\x]<2.9$~MeV at the 90\% C.L.\ is obtained.
This is the most precise measurement of the $\x$ mass and the
most stringent constraint on its width to date, which will
help to refine the parameters of potential models and 
significantly reduce the uncertainties (ca. $\pm50$~MeV) 
of the $D$-wave states predicted by the potential model~\cite{potential2}.


The BESIII collaboration thanks the staff of BEPCII and the IHEP computing center for their strong support. This work is supported in part by National Key R\&D Program of China under Contracts Nos. 2020YFA0406300, 2020YFA0406400; National Natural Science Foundation of China (NSFC) under Contracts Nos. 11975141, 11875115, 11625523, 11635010, 11735014, 11822506, 11835012, 11935015, 11935016, 11935018, 11961141012, 12022510, 12025502, 12035009, 12035013, 12061131003; the Chinese Academy of Sciences (CAS) Large-Scale Scientific Facility Program; Joint Large-Scale Scientific Facility Funds of the NSFC and CAS under Contracts Nos. U1732263, U1832207; CAS Key Research Program of Frontier Sciences under Contract No. QYZDJ-SSW-SLH040; 100 Talents Program of CAS; INPAC and Shanghai Key Laboratory for Particle Physics and Cosmology; ERC under Contract No. 758462; European Union Horizon 2020 research and innovation programme under Contract No. Marie Sklodowska-Curie grant agreement No 894790; German Research Foundation DFG under Contracts Nos. 443159800, Collaborative Research Center CRC 1044, FOR 2359, GRK 214; Istituto Nazionale di Fisica Nucleare, Italy; Ministry of Development of Turkey under Contract No. DPT2006K-120470; National Science and Technology fund; Olle Engkvist Foundation under Contract No. 200-0605; STFC (United Kingdom); The Knut and Alice Wallenberg Foundation (Sweden) under Contract No. 2016.0157; The Royal Society, UK under Contracts Nos. DH140054, DH160214; The Swedish Research Council; U. S. Department of Energy under Contracts Nos. DE-FG02-05ER41374, DE-SC-0012069.


\clearpage
\section*{appendix}

\section{Numerical results of $\sigma[\EE\to\pp\x]\cdot\mathcal{B}[\x\to\gamma\chico]$}

\begin{table}[H]
\begin{center}
\caption{The measured cross section $\sigma[\EE\to\pp\x]$ times the branching ratio $\mathcal{B}[\x\to\gamma\chico]$ 
at different c.m. energies. Here the uncertainties are statistical only.} 
\label{X-sec}
\begin{tabular}{ccccccc}
  \hline\hline
  $\sqrt{s}$~(GeV) & $\mathcal{L}_{\rm int}(\rm pb^{-1})$ & $N^{\rm sig}$ & $\epsilon$ & $1+\delta$ & $\sigma\cdot\mathcal{B}$~(pb) \\
  \hline
  4.2263 & 1056.4 &$1.7^{+2.5}_{-1.7}$ & 0.311 & 0.737 & $0.17^{+0.28}_{-0.18}$ \\
  4.2580 & 828.4 &$1.2^{+2.1}_{-1.3}$ & 0.336 & 0.741 & $0.15^{+0.25}_{-0.16}$ \\
  4.2879 & 502.4 &$0.7^{+1.8}_{-1.0}$ & 0.334 & 0.743 & $0.13^{+0.35}_{-0.19}$ \\
  4.3121 & 501.2 &$0.6^{+1.9}_{-1.0}$ & 0.343 & 0.743 & $0.11^{+0.36}_{-0.20}$ \\
  4.3374 & 505.0 &$3.4^{+2.5}_{-1.8}$ & 0.356 & 0.742 & $0.63^{+0.47}_{-0.34}$ \\
  4.3583 & 543.9 &$6.7^{+3.2}_{-2.5}$ & 0.357 & 0.744 & $1.13^{+0.54}_{-0.42}$ \\
  4.3774 & 522.7 &$8.3^{+3.7}_{-3.0}$ & 0.338 & 0.750 & $1.54^{+0.68}_{-0.55}$ \\
  4.3965 & 507.8 &$12.3^{+4.2}_{-3.5}$ & 0.318 & 0.767 & $2.42^{+0.83}_{-0.69}$ \\
  4.4156 & 1043.9 &$14.2^{+5.2}_{-4.5}$ & 0.310 & 0.798 & $1.35^{+0.49}_{-0.42}$ \\
  4.4362 & 569.9 &$12.5^{+4.3}_{-3.6}$ & 0.323 & 0.841 & $1.98^{+0.67}_{-0.56}$ \\
  4.4671 & 111.1 &$5.3^{+2.9}_{-2.2}$ & 0.332 & 0.910 & $3.85^{+2.09}_{-1.60}$ \\
  4.5271 & 112.1 &$0.0^{+1.6}_{-0.0}$ & 0.320 & 1.017 & $0.00^{+1.07}_{-0.00}$ \\
  4.5745 & 48.9 &$2.0^{+1.8}_{-1.1}$ & 0.307 & 1.053 & $3.02^{+2.82}_{-1.77}$ \\
  4.5995 & 586.9 &$2.1^{+2.5}_{-1.7}$ & 0.318 & 1.014 & $0.27^{+0.32}_{-0.22}$ \\
  4.6120 & 102.5 &$1.5^{+1.9}_{-1.2}$ & 0.328 & 0.960 & $1.12^{+1.45}_{-0.93}$ \\
  4.6278 & 511.1 &$7.0^{+3.8}_{-3.0}$ & 0.348 & 0.860 & $1.12^{+0.60}_{-0.48}$ \\
  4.6408 & 541.4 &$10.0^{+3.9}_{-3.2}$ & 0.371 & 0.783 & $1.56^{+0.60}_{-0.50}$ \\
  4.6613 & 523.6 &$14.3^{+4.5}_{-3.8}$ & 0.384 & 0.796 & $2.18^{+0.69}_{-0.58}$ \\
  4.6811 & 1631.7 &$22.2^{+6.0}_{-5.2}$ & 0.364 & 0.943 & $0.97^{+0.26}_{-0.23}$ \\
  4.6984 & 526.2 &$6.2^{+3.5}_{-2.8}$ & 0.340 & 1.042 & $0.81^{+0.46}_{-0.37}$ \\
  \hline\hline
\end{tabular}
\end{center}
\end{table}

\section{systematic error of resonance parameters}

\begin{table}
\begin{center}
\caption{The systematic uncertainties for the resonance parameters.
$M[R_i]$ and $\Gamma_{\rm tot}[R_i]$ represent the mass (in MeV/$c^2$) and total width (in MeV) of resonance $R_i$, respectively; 
$\Gamma_{\rm e^+e^-} \mathcal{B}_{1}^{R_i} \mathcal{B}_{2}$ is the product of the $\EE$ partial width (in eV/$c^2$) 
and branching fraction of $R_i\to\pp\x\to\pp\gamma\chico$ ($i=1$, $2$). 
The parameter $\phi$ (in degrees) is the relative phase between the two resonances,
and the values in the brackets are the corresponding systematic uncertainties
for the second solution of the two-BW fit.} 
\label{res-par-sys}
\begin{tabular}{ccccc}
\hline\hline
Parameters & $\sqrt{s}$ & $\sigma\cdot\mathcal{B}$ & Fit model & Sum \\
\hline
$M[R_1]$  & 3.9 & -- & 2.2 & 4.5 \\
$\Gamma_{\rm tot}[R_1]$ & 1.6 & -- & 1.6 & 2.3 \\
$\Gamma_{\rm e^+e^-} \mathcal{B}_{1}^{R_1} \mathcal{B}_{2}$ & 0.01 (0.01) & 0.03 (0.03) & 0.01 (0.01) & 0.03 (0.03) \\
$M[R_2]$  & 0.7 & -- & 0.4 & 0.8 \\
$\Gamma_{\rm tot}[R_2]$ & 0.4 & -- & 4.1 & 4.1 \\
$\Gamma_{\rm e^+e^-} \mathcal{B}_{1}^{R_2} \mathcal{B}_{2}$ & 0.01 (0.01) & 0.02 (0.01) & 0.01 (0.01) & 0.02 (0.01) \\
$\phi$ & 0.3 (1.8) & -- & 3.2 (5.4) & 3.2 (5.7) \\
\hline
$M[R]$  & 3.2 & -- & 1.3 & 3.5 \\
$\Gamma_{\rm tot}[R]$ & 1.7 & -- & 12.2 & 12.3 \\
$\Gamma_{\rm e^+e^-} \mathcal{B}_{1}^{R} \mathcal{B}_{2}$ & 0.01 & 0.05 & 0.02 & 0.05 \\
\hline\hline
\end{tabular}
\end{center}
\end{table}

\section{Results of $\mathcal{B}[\x\to\gamma\chict]$}

\begin{figure}
\begin{center}
\includegraphics[height=2in]{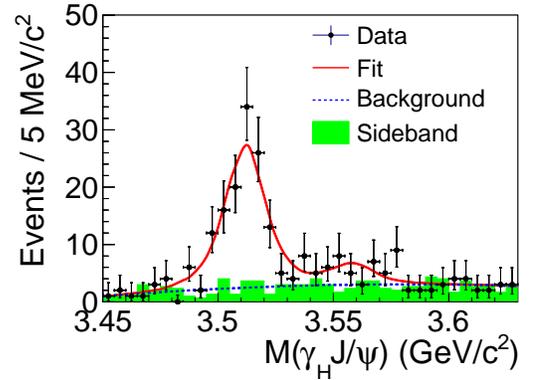}
\caption{Result of the fit to the $M(\gamma_{H}\jpsi)$ distribution
for the events in the $\x$ signal region ($3.815<M^{\rm recoil}(\pp)<3.835$~GeV/$c^2$). 
Dots with error bars are data, the red solid curve is the total fit, the blue
dashed curve is background, and the green shaded
histogram is the background estimated from $\x$ sideband events.
}
\label{chic2}
\end{center}
\end{figure}

For the $\x\to\gamma\chict$ decay, we study the $M(\gamma_{H}\jpsi)$ distribution
by requiring $3.815<M^{\rm recoil}(\pp)<3.835$~GeV/$c^2$ to select $\x$ signal candidates.
In order to estimate non-$\x$ background, we also define a sideband region 
as $3.74<M^{\rm recoil}(\pp)<3.78$~GeV/$c^2$. Figure~\ref{chic2} shows
the $M(\gamma_{H}\jpsi)$ distribution, where no significant $\chict$ signal is seen. 
A fit with $\chico$ and $\chict$ signal shapes determined from MC simulation 
as the signal PDF, 
and a second-order polynomial as the background is used to extract the relative decay rate of
$R=\frac{\mathcal{B}[\x\to\gamma\chict]}{\mathcal{B}[\x\to\gamma\chico]}=0.33\pm0.12$.
Since the $\chict$ signal is not significant (the statistical significance is only $2.0\sigma$), 
an upper limit of $R<0.51$ 
at the $90\%$ C.L. is given, taking into account the systematic uncertainty.

\section{Scattering angle distribution}

\begin{figure}
\begin{center}
\includegraphics[height=2in]{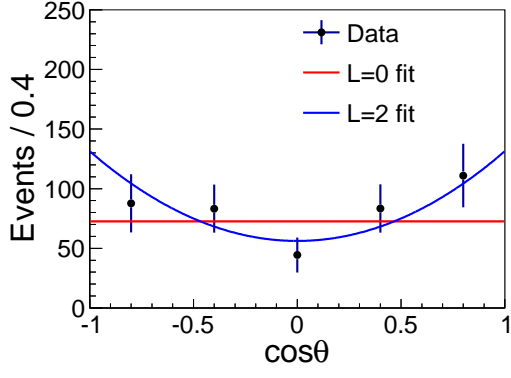}
\caption{Scattering angle distribution for $\x$ events in $\EE$ CM frame (after efficiency correction).
Dots with error bars are data, the red and blue curves are from the $L=0$ and $L=2$ fits, respectively.}
\label{angle}
\end{center}
\end{figure}

The $\pp$ system in the $\EE\to\pp\x$ process is expected to be dominated by $S$-wave,
such as $f_0(500)$. According to spin-parity conservation, the orbital angular
momentum $L$ between $\pp$ and $\x$ is therefore $2$.
With helicity amplitude calculations, the scattering angle distribution of $\x$ 
is ($1+\cos^2\theta$), where $\theta$ is the polar angle of $\x$ in the $\EE$ c.m.\ frame. 
Figure~\ref{angle} shows the $\cos\theta$
distribution of the selected $\EE\to\pp\x$ signal candidates after efficiency correction. 
We perform fits to the angular distribution with an $L=0$ PDF (flat) 
and an $L=2$ PDF ($1+\alpha\cos^2\theta$, where $\alpha=1.3\pm 0.8$ is obtained from the fit).
A $\chi^2$-test for the $L=2$ fit yields $\chi^2/ndf=2.3/3=0.8$, 
which is better than that of the $L=0$ fit ($\chi^2/ndf=6.8/4=1.7$).

\end{document}